\documentstyle[aps,prl,epsfig,twocolumn]{revtex}

\def\be{\begin{equation}}
\def\ee{\end{equation}}
\def\bea{\begin{eqnarray}}
\def\eea{\end{eqnarray}}
\def\bma{\begin{mathletters}}
\def\ema{\end{mathletters}}
\def\C{\hbox{$\mit I$\kern-.6em$\mit C$}}

\begin{document}
\draft

\title{Transmitting One Qubit Can Increase One Ebit Between Two Parties at Most}

\author{Yi-Xin Chen and Dong Yang}

\address{Zhejiang Institute of Modern Physics and
Department of Physics, Zhejiang University, Hangzhou 310027, P.R. China} 

\date{\today}

\maketitle

\begin{abstract}
Entanglement shared between two seperated parties could not be increased without transmitting quantum system. We suggest the project to gain entanglement shared between Alice and Bob by transmitting quantum system and a new scheme to achieve efficient rate of entanglement. It is proven that under any local operations and communicating classically, to transmit one qubit through an ideal or a noisy quantum channel can increase no more than one ebit between two parties. Furthermore, the prior nonmaximally transmitted entanglement could not be improved by subsequently transmitted qubit. Although our proof is given in the measure of formation of entanglement, we believe that the above conclusion is also hold independently of the measure of entanglement. 
\end{abstract}


\narrowtext

Quantum correlation, known also as entanglement, is one of the most fascinating properties of composite quantum system. Entanglement is related with many quantum phenomena, such as quantum information \cite{Schumacher,Barnum,Adami}, quantum computation \cite{Barenco}, quantum error-correction \cite{Bennett1}, in particular teleportation \cite{Bennett2}. Recently, many contributions were made to quantifying entanglement of mixed state \cite{Vedral}, purification of mixed state \cite{Bennett1,Bennett3}, multiparticle entanglement \cite{GHZ,Vidal}. Most of the papers suppose entanglement shared beforehand between two parties, traditionally named Alice and Bob. In this Letter, we consider the step before entanglement is utilized to carry out some tasks, that is, how they achieve entanglement shared between the two seperated parties. Bennett et al. \cite{Bennett4} has pointed out that one qubit can be used to create one ebit of entanglement. Vedral et al. \cite{Vedral} proved that two parties can not, no matter how small probability, by local operations and communicating classically turn a disentangled state $\sigma_{AB}$ into an entangled state and that by local operation and classical communication alone, they cannot increase the total amount of entanglement which they share. So the only way to create and increase entanglement between two parties is to transmit quantum system which carries quantum information on entanglement. Now, our problems arise: What is the efficiency of transmitting qubit to obtain entanglement? How much entanglement could be transmitted by transmission of one qubit? If the prior qubit transmitted with non-maximal entanglement is achieved, could the subsequent transmission of qubit compensate for its predecessor's slip? If it is yes, this means that the latter transmitted could increase more than one ebit \cite{Bennett4}.

Suppose Alice and Bob share no entanglement beforehand. The only way to achieve shared entanglement is to transmit quantum system from Alice to Bob. Here, we suppose the basic quantum system lies in two dimensional Hilbert space which does not lose any generality. Two dimensional quantum system is also named quantum bit abbreviated qubit following Schumacher \cite{Schumacher}. Alice has four parties denoted as A, B, C, D, in which she wants to transmit two of them C and D to establish entanglement with Bob for some tasks later. First we suppose the quantum channel is ideal and then generalize it to be noisy.

Theorem 1: If the state of composite system ABCD remains pure from beginning to end, whatever unitary operations are performed by Alice, sending subsystem D to Bob can create no more than one ebit between them. Once Alice has transmitted D with nonmaximal entanglement between subsystum ABC and D, she could not correct it as maximal entanglement through transmitting C and obtain two ebits shared at the end.

Proof: 
We denote $|\phi_{ABCD}>$ as the state of the composite system ABCD, $E(|\phi_{ABC\sim D}>)$ as the amount of entanglement between subsystum ABC and D in state $|\phi_{ABCD}>$, and $E(|\phi_{AB\sim CD}>)$ as the amount of entanglement between subsystum AB and CD in the same state $|\phi_{ABCD}>$.
 
Step 1, Alice can use unitary operation $U^{ABCD}$ to manipulate arbitary pure state $|\phi_{ABCD}>$ from product state $|0_{A}>|0_{B}>|0_{C}>|0_{D}>$. 
According to Schmidt decomposition \cite{Peres}, 
\be 
|\phi_{ABCD}>=a|\phi_{ABC}>|\phi_{D}>+b|\phi_{ABC}^{\perp}>|\phi_{D}^{\perp}> ,
\ee
in which 
\bea
<\phi_{ABC}|\phi_{ABC}^{\perp}> &=& 0 ,\nonumber\\
<\phi_{D}|\phi_{D}^{\perp}> &=& 0 .\nonumber
\eea

Step 2, Alice transmits D to Bob through an ideal channel. Now the entanglement shared between them is, 
\be
E(|\phi_{ABC\sim D}>)
=S(\rho_{ABC})=S(\rho_{D}) ,
\ee
where $S(\rho)$ is von Neumann entropy \cite{Neumann}, $S(\rho)= -Tr\rho \log \rho$,  $\rho_{ABC}$ and $\rho_{D}$ are the marginal density operators of state $|\phi_{ABCD}>$ which are obtained by tracing D and ABC out respectively. The amount of entanglement of pure state is measured by the von Neumann entropy of its marginal density operator seen by either party\cite {Bennett4}.

Step 3, Now the allowed operation performed by Alice is $U^{ABC}$, one by Bob is $U^{D}$, then
\bea 
&U&^{ABC}\otimes U^{D}|\phi_{ABCD}> \nonumber \\
&=& U^{ABC}\otimes U^{D}(a|\phi_{ABC}>|\phi_{D}>+b|\phi_{ABC}^{\perp}>|\phi_{D}^{\perp}>) \nonumber \\
&=& a|\psi_{ABC}>|\psi_{D}>+b|\psi_{ABC}^{\perp}>|\psi_{D}^{\perp}> .
\eea
From the above equation, we can see clearly that any local unitary operations performed by Alice and Bob could not change the coefficients a and b which are determined by the operation $U^{ABCD}$ performed by Alice. Coefficients a and b represent the entanglement between Alice and Bob for the transmission of subsysten D. This property confines the consecutive correction of entanglement and limit the efficiency of entanglement transmission as will be showed next step. Another important point is that suppose U is the set of the unitary operations satisfying 
\bea
U &=& \{U^{ABCD}|U^{ABCD}|0_{A}>|0_{B}>|0_{C}>|0_{D}> \nonumber \\
 &=& a|\phi_{ABC}>|\phi_{D}>+b|\phi_{ABC}^{\perp}>|\phi_{D}^{\perp}>, a, b~~ fixed\} ,
\eea
all the state of the form could be realized by $U^{ABC}\otimes U^{D}$ performed on one of the state of the above form. That is
\bea
\{U^{ABCD}|0_{A}>|0_{B}>|0_{C}>|0_{D}>|U^{ABCD}\in U\}= \nonumber \\
\{U^{ABC}\otimes U^{D}(a|\phi_{ABC}>|\phi_{D}>+b|\phi_{ABC}^{\perp}>|\phi_{D}^{\perp}>)\} .
\eea
We consider this sheds lights on relation between the collective unitary operation and product operation.

Now let us give some comments on Step 3. From above discussion, we know that operations performed by Alice and Bob could not change a and b which means they could not change entanglement between them, more precisely could not increase entanglement. Ideally, one may hope that Alice and Bob can perform some optimal operations on their respective subsystem so that after transmitting subsystem C, they could obtain maximal entanglement between AB and CD which means that entanglement increases more than one ebit after subsystem C is transmitted. Our answer is negative. We have showed that after any operation $U^{ABC}\otimes U^{D}$, the state is of the form $a|\psi_{ABC}>|\psi_{D}>+b|\psi_{ABC}^{\perp}>|\psi_{D}^{\perp}>$. So the problem converts to a new one: What is the maximum of entanglement between subsystum AB and CD in the state of the above form? After finding the optimal form, Alice and Bob change the state to that form.

Step 4, Alice sends subsystem C to Bob. Now they share entanglement between AB and CD. They can not increase entanglement by any local operations and classical communications. We can describle the composite operation of the above steps in the following process, 
\\~~~\\ 
$|0_{A}>|0_{B}>|0_{C}>|0_{D}>~~~  
 \underline{1:U^{ABCD}} \mapsto
|\phi_{ABCD}> 
\\~~~\\
 \underline{2: \roarrow{D}} \mapsto  
|\phi_{ABC\sim D}> ~~~ 
 \underline{3:U^{ABC}\otimes U^{D}} \mapsto
|\psi_{ABC\sim D}> ~~~ 
\\~~~\\
 \underline{4: \roarrow{C}} \mapsto 
|\psi_{AB\sim CD}>$ .
\\~~~\\
The variance of entanglement shared between Alice and Bob can be read from the following expressions.
\\After Step 1, $E_{1}=0$ .
\\After Step 2,
\be
E_{2} = E(|\phi_{ABC\sim D}>) 
=S(\rho_{ABC}^{2})=S(\rho_{D}^{2}) ,
\ee
where 
\bea
\rho_{ABC}^{2} = Tr_{D}|\phi_{ABCD}><\phi_{ABCD}| \nonumber ,
\\\rho_{D}^{2} = Tr_{ABC}|\phi_{ABCD}><\phi_{ABCD}| \nonumber .
\eea
After Step 3,
\be
E_{3} = E(|\psi_{ABC\sim D}>)=S(\rho_{ABC}^{3})=S(\rho_{D}^{3}) ,
\ee
where
\bea
\rho_{ABC}^{3} &=& Tr_{D}|\psi_{ABCD}><\psi_{ABCD}| ,\nonumber
\\\rho_{D}^{3} &=& Tr_{ABC}|\psi_{ABCD}><\psi_{ABCD}| .\nonumber
\eea
According to the property of $S(\rho)$ under unitary operations, 
\bea
S(\rho_{ABC}^{3}) &=& S(\rho_{ABC}^{2}) ,\nonumber
\\S(\rho_{D}^{3}) &=& S(\rho_{D}^{2}) ,\nonumber
\eea
so 
\be
E_{3} = E_{2} .
\ee
After Step 4, 
\be
E_{4}=E(|\psi_{AB\sim CD}>)
=S(\rho_{AB}^{4})=S(\rho_{CD}^{4}) ,
\ee
where
\bea
\rho_{AB}^{4} = Tr_{CD}(|\psi_{ABCD}><\psi_{ABCD}|) ,\nonumber
\\\rho_{CD}^{4} = Tr_{AB}(|\psi_{ABCD}><\psi_{ABCD}|) .\nonumber
\eea
The increasion of entanglement of transmission D,
\be
E_{2}-E_{1}=S(\rho_{D}^{2})\leq 1 .
\ee
The increasion of entanglement of transmission C,
\be 
E_{4}-E_{3}=S(\rho_{AB}^{4})-S(\rho_{ABC}^{3}) .
\ee
According to the inequality of entropy \cite{Wehrl},
\be
|S(\rho_{AB})-S(\rho_{C})|\leq S(\rho_{ABC})\leq S(\rho_{AB})+S(\rho_{C}) ,
\ee
and $\rho_{AB}^{4} = Tr_{C}\rho_{ABC}^{3}$, we know
\bea
E_{4}-E_{3} &\leq& S(\rho_{C}^{4}) ,
\\E_{4} \leq E_{2}+S(\rho_{C}^{4})&\leq& E_{2}+1\leq 2 .
\eea

The same conclusion can be proved from Bob's part. Thus we have completed the proof of theorem one. From the above, we mention that it is also hold in the case of unequal subsystems between Alice and Bob and easily generalized to be in the case of more than two dimensionality. Furthermore, the equality is hold if and only if subsystums C and D are respectively maximally entangled with subsystum AB and they themselves are independent of each other.

Theorem 2: Theorem 1 is also hold if the state of composite system ABCD is mixed.

We prove it in the measure of formation of entanglement and argue for it independent of measures of entanglement. The similar process in the proof theorem 1 is the following, where the initial density matrix is mixed. 
\\~~~\\
$\rho_{ABCD} 
\underline{1:U^{ABCD}} \mapsto
\rho_{ABCD}^{1} 
\underline{2:\roarrow{D}} \mapsto   
\rho_{ABC\sim D}^{2}=\rho_{ABCD}^{1}$ 
\\~\\
$\underline{3:U^{ABC}\otimes U^{D}}\mapsto  
\rho_{ABC\sim D}^{3} 
\underline{4:\roarrow{C}}\mapsto 
\rho_{AB\sim CD}^{4}=\rho_{ABC\sim D}^{3}$.
\\~~~\\
The amount of formation of entanglement $\sigma$ is defined as \cite{Bennett4},
\be
E_{f}(\sigma_{AB})=min_{\sigma_{AB}=\sum_{i} p_{i}|\psi_{AB}>_{ii}<\psi_{AB}|}\sum_{i} p_{i}E(|\psi_{AB}>_{i}) .
\ee

Proof:
 Suppose $\{|\phi_{ABCD}>_{i},p_{i}\}$ is the pure state ensemble which minimizes 
$\sum_{i} p_{i} E(|\phi_{ABC\sim D}>_{i})$.
\\After Step 1, $E_{1}=0$ .
\\After Step 2,
\be
E_{2}=E_{f}(\rho_{ABC\sim D}^{2}) .
\ee
After Step 3,
\bea
&\rho&_{ABCD}^{3} = U^{ABC} \otimes U^{D} \rho_{ABCD}^{2} U^{ABC\dagger} \otimes U^{D\dagger} \nonumber \\
&=& U^{ABC} \otimes U^{D} \sum_{i} p_{i}|\phi_{ABCD}>_{ii}<\phi_{ABCD}| U^{ABC\dagger} \otimes U^{D\dagger} \nonumber \\
&=& \sum_{i} p_{i} |\psi_{ABCD}>_{ii}<\psi_{ABCD}| .
\eea
So $\{|\psi_{ABCD}>_{i},p_{i}\}$ is an pure states ensemble realizing $\rho_{ABCD}^{3}$, we know 
\be
E_{f}(\rho_{AB\sim CD}^{3}) \leq \sum_{i}p_{i}E(|\psi_{AB\sim CD}>_{i}) ,\nonumber
\ee
\be
E_{3}=E_{f}(\rho_{ABC\sim D}^{3})=E_{f}(\rho_{ABC\sim D}^{2}) .
\ee
After Step 4, $\rho_{ABCD}^{4} = \rho_{ABCD}^{3}$ ,
\be
E_{4} = E_{f}(\rho_{AB\sim CD}^{4})=E_{f}(\rho_{AB\sim CD}^{3}) .
\ee
According to Theorem One, to any pure state, the inequality  
\be
E(|\psi_{AB\sim CD}>_{i}) \leq E(|\phi_{ABC\sim D}>_{i})+1 ,\nonumber
\ee
is hold, so
\be
E_{4} \leq  \sum_{i}p_{i}(E(|\phi_{ABC\sim D}>_{i})+1)=E_{2}+1 .
\ee

We have established theorem 2. Though we prove theorem 2 in the measure of formation, we conjecture it is hold independent of the measures of entanglement. We argue that the more one qubit entangled with other quantum system, the more mixed the qubit is. However, the maximally mixed state of one qubit is $1/2I$ which can only carry one ebit at most.

Theorem 3: Theorem 2 is hold through noisy channel.

We adopt Schumacher's discription about noisy channel ${\cal E}$  \cite{Schumacher}, suppose an environment E is initially in the pure state $|0_{E}>$, after quantum system Q in initial state $\rho$ is subjected to dynamical process ${\cal E}$, we could have  
\be 
{\cal E}(\rho)=Tr_{E}U^{QE}(\rho \otimes |0_{E}><0_{E}|)U^{QE\dagger} .\nonumber
\ee
The dynamical process is describled by
\\~\\
$\rho_{ABCD}~~~ 
\underline{1:U^{ABCD}} \mapsto
\rho_{ABCD}^{1}~~~   
\underline{2:\roarrow{{\cal E}(D)}} \mapsto   
\rho_{ABC\sim D}^{2}~~~ 
\\~\\
\underline{3:U^{ABC}\otimes U^{D}}  \mapsto   
\rho_{ABC\sim D}^{3}~~~ 
\underline{4:\roarrow{{\cal E}(C)}} \mapsto
\rho_{AB\sim CD}^{4}$.
\\~\\

Proof:
\be
E_{f}(\rho_{AB\sim CD}^{4}) \leq E_{f}(\rho_{AB\sim CD}^{3}) 
\leq E_{f}(\rho_{ABC\sim D}^{2})+1
\ee

Bennett et al. \cite{Bennett1} have proved that entanglement of formation is noncreasing under local operations and classical communication. The left inequality results from the fact that local operations can not increase the entanglement and the noisy channel may be regarded as the operation performed by Bob. Hence it is clear that there exists theorem 3.

Theorem 4: Theorem 3 is hold if Alice and Bob are allowed to perform general unitary operation and classical communication. The dynamical process is  
\\~\\
$\rho_{ABCD}~~~  
\underline{1:U^{ABCD}} \mapsto 
\rho_{ABCD}^{1}~~~   
\underline{2:\roarrow{{\cal E}(D)}} \mapsto  
\rho_{ABC\sim D}^{2}~~~ 
\\~\\
\underline{3:\sum_{i}p_{i}U^{ABC}_{i}\otimes U^{D}_{i}} \mapsto     
\rho_{ABC\sim D}^{3}~~~ 
\underline{4:\roarrow{{\cal E}(C)}} \mapsto
\rho_{AB\sim CD}^{4}$.
\\~\\

Proof:
\bea
E_{f}(\rho_{AB\sim CD}^{4}) &\leq& E_{f}(\rho_{AB\sim CD}^{3}) \nonumber \\
&\leq& E_{f}(\rho_{AB\sim CD}^{2}) \nonumber \\
&\leq& E_{f}(\rho_{ABC\sim D}^{2})+1 .
\eea

The reason of the first inequality is the same as Theorem 3. The second inequality is hold because the general unitary operations and classical communication performed by the two parties could be represented by operator $\sum_{i} q_{i}U_{i}^{ABC} \otimes U_{i}^{D}$ while the inequality is hold for each item. That ends the proof.

In this Letter, we have discussed the bound efficiency of achieving shared entanglement and concluded that one qubit can carry on one ebit at most. The given project of transmitting quantum system can be used not only to build up entanglement shared between two seperated parties but also to measure the amount of transmitting quantum information by locally  determining entangelment and classically communicating between the parties. From a new point of view, the nature of sharing information is to share correlation between two parties. The classical information theory is in making classical correlation, while the quantum counterpart is in making quantum correlation which is the most intriguing property distinguished from classical one. In classical information theory, the binary signal is the carrier of information and its capacity of information is one bit. In quantum information theory, the qubit is the carrier of quantum information and its capacity of quantum information is one ebit, In the quantum case, we would define as the amount of entanglement shared between two parties and would discuss it parellel to classical one in future papers.

D. Yang thanks S.J. Gu for helpful discussion.
The work is supported by
the NNSF of China (Grant No.19875041), the Special 
NSF of Zhejiang Province (Grant No.RC98022) and Cao Guang-Biao Foundation in 
Zhejiang University.



\begin{references}

\bibitem{Barenco}
A. Barenco, Contemporary Physics {\bf 37}, 375 (1996).

\bibitem{Bennett1}
C.H. Bennett, D.P. DiVincenzo, J.A. Smolin and W.K. Wootters, Phys. Rev. A {\bf 54}, 3824 (1996).

\bibitem{Bennett2}
C.H. Bennet, G. Brassard, C. Crepeau, R. Jozsa, A. Peres and W.K. Wootters, 
Phys. Rev. Lett {\bf 70}, 1895 (1993).

\bibitem{Schumacher} 
B.W. Schumacher, Phys. Rev. A {\bf 54}, 2614 (1996);
B.W. Schumacher and M.A. Nielson, {\it ibid}. {\bf 54}, 2629 (1996).

\bibitem{Barnum}
H. Barnum, M.A. Nielson and B.W. Schumacher, Phys. Rev. A {\bf 57}, 4153 (1998);
H. Barnum, E. Knill and M.A. Nielson, quant-ph/9809010.

\bibitem{Adami}
C. Adami and N.J. Cerf, Phys. Rev. A {\bf 56}, 3470 (1997);
N.J. Cerf and C. Adami, Phys. Rev. Lett. {\bf 79}, 5194 (1997).

\bibitem{Vedral}
V.Vedral and M.B. Plenio, Phys. Rev. A {\bf 57}, 1619 (1998).
 
\bibitem{Bennett3}
C.H. Bennett, G.Brassard, Sandu Popescu, Benjamin Schumacher, J.A. Smolin and W. K. Wootters, Phys. Rev. Lett. {\bf 76}, 722 (1996).

\bibitem{Bennett4}
C.H. Bennett, H.J. Bernstein, Sandu Popescu, Benjamin Schumacher, Phys. Rev. A {\bf 53}, 2046 (1996).

\bibitem{GHZ}
D.B. Greenber, M.Horhe and A.Zeilinger, Physics Today, {\bf 42}, 22 (1993)

\bibitem{Vidal}
G.Vidal, W.Dur and J.I.Cirac, quant-ph/0004009

\bibitem{Peres}
Asher Peres, Quantum Theory:Concepts and Methods, 
(Kluwer Academic Publishers, 1993)

\bibitem{Neumann}
J. von Neumann, Mathematical Foundations of Quantum Mechanics,
( Princeton, New Jersey, 1955 ). 

\bibitem{Wehrl}
A. Wehrl, Rev. Mod. Phys. {\bf 50}, 221 (1978).

\end{references}
\end{document}